\documentclass[aoas,MSNbibl,nameyear,seceqn,dvips]{arximspdf}
\usepackage{dcolumn}
\usepackage{graphicx}

\doi{10.1214/12-AOAS588} 
\volume{7}
\issue{1}
\pubyear{2013}
\firstpage{81}
\lastpage{101}

\makeatletter

\newcolumntype{d}[1]{D{.}{.}{#1}}

\newcommand{\rrVert}{\Vert}
\newcommand{\rrvert}{\vert}
\newcommand{\llVert}{\Vert}
\newcommand{\llvert}{\vert}
\makeatother

\begin{document}
\begin{frontmatter}

\title{Geostatistical modeling in the presence of interaction between
the measuring instruments, with an application to the estimation of
spatial market potentials}
\runtitle{Geostatistical modeling under spatial interaction}

\begin{aug}
\author[A]{\fnms{Francesco} \snm{Finazzi}\corref{}\ead[label=e1]{francesco.finazzi@unibg.it}}
\runauthor{F. Finazzi}
\affiliation{University of Bergamo}
\address[A]{Department of Information Technology\\
\quad and Mathematical Methods\\
University of Bergamo\\
Viale Marconi\\
5-24044 Dalmine (BG)\\
Italy\\
\printead{e1}} 
\end{aug}

\received{\smonth{3} \syear{2012}}
\revised{\smonth{5} \syear{2012}}

%
\begin{abstract}
This paper addresses the problem of recovering the spatial market
potential of a retail product from spatially distributed sales data. In
order to tackle the problem in a general way, the concept of spatial
potential is introduced. The potential is concurrently measured at
different spatial locations and the measurements are analyzed in order
to recover the spatial potential. The measuring instruments used to
collect the data interact with each other, that is, the measurement at
a given spatial location is affected by the concurrent measurements at
other locations. An approach based on a novel geostatistical model is
developed. In particular, the model is able to handle both the
measuring instrument interaction and the missing data. A model
estimation procedure based on the expectation--maximization algorithm is
provided as well as standard inferential tools. The model is applied to
the estimation of the spatial market potential of a newspaper for the
city of Bergamo, Italy. The estimated spatial market potential is
eventually analyzed in order to identify the areas with the highest
potential, to identify the areas where it is profitable to open
additional newsstands and to evaluate the newspaper total market volume
of the city.
\end{abstract}

%

\end{frontmatter}

\section{Introduction}\label{sec1}
The market potential of a given retail product is the expected sales
volume when the product is marketed.
The spatial market potential is the spatial distribution of the market
potential over a trading area.
Sales are expected to be high if a store is opened at a spatial
location characterized by a high spatial market
potential, while they are expected to be low if the spatial location
has a low spatial market potential.

With the goal to increase and to maximize the sales volume, a key issue
is how to evaluate the spatial
market potential. In this paper, it is assumed that the product is
already marketed and that the sales
data of spatially distributed stores are available. Thus, the aim is to
estimate the spatial market
potential by considering the sales data, the spatial characteristics of
the trading area and
the spatial interaction between the stores.
The stores interact in the sense that the sales volume of each store is
affected by the presence of all
the others. As a consequence, the spatial market potential cannot be
estimated ignoring the interaction.

For all purposes and intents, the spatial market potential can be
regarded as a spatial
surface, as it is well defined for all the spatial locations of the
trading area. Taking a statistical
perspective, the spatial market potential is considered as a spatially
continuous random field and the estimation
of the spatial market potential is obtained through the estimation of
the realization of the random field.
Although well understood, however, no attempt has ever been made to
address the problem following a geostatistical approach.

The estimation of a spatial market potential is an instance of the more
general problem of
recovering the realization of a spatially continuous random field in
the case of interacting
measuring instruments. The instruments interact in the sense that the
measurement at
a given spatial location is affected by the concurrent measurements at
nearby locations.

A novel model able to handle both the interaction between the measuring
instruments and the missing data
is proposed. A case study is presented, in which the sales data of
spatially distributed
newsstands are used to estimate the spatial market potential of an
economic daily newspaper
for the city of Bergamo, Italy. The aim of the study is threefold: to
identify the areas with
the highest market potential, to identify the areas where it is
profitable to open additional
newsstands and to estimate the total market volume of the city with
respect to the
newspaper considered.

The rest of the paper is organized as follows: Section~\ref{sec2} provides the
background and motivation for this work.
Section~\ref{sec3} introduces a novel geostatistical model for the analysis of
spatial point data in the case of
interaction between the measuring instruments. Model estimation and
inference are discussed in Section~\ref{sec4}.
Section~\ref{sec5} presents the case study while Section~\ref{sec6} provides conclusions.
The technical aspects related to
the model estimation are reported in the \hyperref[app]{Appendices}.

\section{Background}\label{sec2}
In this section the spatial market potential estimation problem is
discussed in terms of both
the current state of the art and the available statistical methods. It
turns out that the
estimation of a spatial market potential from sales data received
little attention
in the past and that the classic geostatistical approach cannot be
adopted to solve the problem.

\subsection{Spatial market potential estimation}\label{sec21}
The problem of estimating the market potential of a retail product is not
new in the literature. The state of the art is represented by the
so-called\vadjust{\goodbreak}
spatial interaction models which describe the market potential in terms
of flows between a set of origins (the customers) and a set of
destinations (the stores).
The interested reader is referred to the seminal papers of Reilly and Huff
[\citet{reilly1931}, \citet{Huff1964}] and to the more recent literature
[see, e.g., \citet{Davis2006}, \citet{cliquet2006},
\citet{Grange2009} and \citet{Manfred2011}].
The spatial interaction models focus on the utility that consumers
obtain from buying a retail product
at a specific store. The utility is often a function of the attributes
of the product, the attributes of
the store and some attributes of separation such as the geographic
distance, the transport cost and the transport time.

The main drawback of the spatial interaction models is that the spatial
market potential is not explicitly
modeled as a regionalized variable and it is defined only at the spatial
location of the stores. This is in contrast with the concept of spatial
market potential adopted in this
paper, which is supposed to exist beyond the existence of the stores.
Indeed, the market potential
of a product at a given location in space can also be considered as the
willingness of the
consumers to reach that specific location in order to buy the product.
The attributes of the
store (including the price at which the store sells the product) may
affect the way the spatial
market potential is observed, but the spatial market potential is not
driven by the stores.

The spatial interaction models literature also lacks of methods for
estimating the model
output uncertainty. This represents a critical issue, as, in practical
applications, the available
data are usually limited in number and the reliability of the model
output must be provided.

Modeling the market potential as a regionalized variable, and, in
particular, as a spatially continuous random field,
allows to answer new and interesting questions. Denoting $q(\mathbf
{s})$ the spatial
market potential at the generic spatial location $\mathbf{s}$, the
company that owns the
stores may be interested in estimating $q(\mathbf{s})$ for each point
of the trading area $\mathcal{D}$.
For instance, the company may want to locate the maxima of $q(\mathbf
{s})$ to be sure
that it has a store near that location. If, instead, the company wants
to open a new store,
then it may want to evaluate the market potential conditioned on the
presence of
the actual stores. Moreover, the company may want to pursue both of the
goals even if the
sales data of some stores are missing. In this sense, when a latent
spatial market potential has
to be assessed and the uncertainty information must be provided, the
geostatistical approach seems
to be more appropriate than any approach based on the spatial
interaction models.

\subsection{Geostatistical modeling}\label{sec22}

The problem of estimating a spatially continuous random field from
measurements collected at a finite number of locations in space is
usually solved by considering geostatistical models and kriging
techniques [see \citet{cressiewikle2011}].

The simplest geostatistical model described in \citet{Diggle2007},
for instance, assumes\vadjust{\goodbreak} an underlying stationary Gaussian random field
$w(\mathbf{s%
})$ and observation $y(\mathbf{s}_{i})$ which are realizations of
conditionally mutually
independent random variables $Y ( \mathbf{s}_{i} ) $
conditionally normally
distributed with mean $E ( Y ( \mathbf{s}_{i} )
\mid w ( \cdot )  )=w ( \mathbf{s}_{i}
) $ and variance $\sigma^{2}_{\varepsilon}$. In the spatial market
potential case, however, due
to the interaction between the stores, the conditional mutual
independence of the random
variables $Y ( \mathbf{s}_{i} ) $ is not met and, in
general, $%
E ( Y ( \mathbf{s}_{i} ) \mid q ( \cdot )
)
\neq q ( \mathbf{s}_{i} )$.
A geostatistical approach for spatial interaction data can be found in
\citet{Banerjee2000},
though the interaction is defined in terms of flows between
destinations and origins
(cf. the previous paragraph) and the approach is not suitable for the
problem addressed in
this paper, where a ``diffuse'' origin is considered.

\section{The geostatistical potential model}\label{secGIM}\label{sec3}

\subsection{Introduction}\label{parnewapproach}\label{sec31}

Before developing a suitable geostatistical model, the problem at hand is
restated and generalized in the following way.

Let $q(\mathbf{s})$ be a spatial random field defined over the region
of space
$\mathcal{D}\subset\mathbb{R}^{2}$. The random field $q(\mathbf
{s})$ is called here \textit{potential}, though the term does not
refer to any particular property of the field. The random field is
concurrently measured at the set of spatial locations
$\mathcal{S}= \{\mathbf{s}_{1},\ldots,\mathbf{s}_{N} \}$ and
the observations
$\mathbf{y} (\mathcal{S} ) $ are collected (possibly with
missing data).
The concurrency of the measurements is a key aspect in the sense that, in
general, the observations $\tilde{\mathbf{y}} ( \mathcal
{S} ) $
collected in the case of nonconcurrent measurements differ from
$\mathbf{y}%
( \mathcal{S} ) $.

The observations $\mathbf{y} ( \mathcal{S} ) $ are
supposed to be
realizations of random variables $Y ( \mathbf{s}_{i} ) $
conditionally normally
distributed with conditional mean%
\[
E \bigl( Y ( \mathbf{s}_{i} ) \mid q ( \cdot ), \mathcal{S}%
\bigr) =h \bigl( q ( \mathbf{s}_{i} );\mathcal{S} \bigr);\qquad
\mathbf{s}_{i}\in\mathcal{S},i=1,\ldots,N,
\]
and conditional variance $\sigma^{2}_{\varepsilon}$, where $h$ is a
function modeling the interaction
between the measuring instruments.

The interaction between the measuring instruments is said to be
an \textit{absorption interaction} if, for each $\mathbf{s}\in
\mathcal{D}$,
$h ( q ( \mathbf{s}_{i} );\mathcal{S} ) <
q ( \mathbf{s} )$.
The interaction is such that, for each $\mathbf{s}%
_{1}\in\mathcal{D}$, $h ( q ( \mathbf{s}_{1}
);\mathcal{S}%
) \equiv q ( \mathbf{s}_{1} ) $ iff $\mathcal
{S=} \{
\mathbf{s}_{1} \} $, that is, the conditional mean of $Y (
\mathbf{s%
}_{1} ) $ is equal to $q ( \mathbf{s}_{1} ) $ if
$\mathbf{s}%
_{1}$ is the only spatial location of $\mathcal{D}$ where $q$ is measured.
Ultimately, it can be stated that the act of measuring the potential
$q$ at
a given $\mathbf{s}$ alters the (concurrent) measurements at other spatial
locations.

At this point, the following distinction can be made: the potential
$q(\mathbf{s})$
is the expected observed value when $q$ is measured only at
the spatial location $\mathbf{s}\in\mathcal{D}$. On the other hand,
the \textit{conditional potential} $q(\mathbf{s};\mathcal{S})$ is the expected
observed value when $q$ is measured at the spatial
location $\mathbf{s}\in\mathcal{D}$ given that it is concurrently measured
at the set of locations $\mathcal{S}= \{ \mathbf
{s}_{1},\ldots,\mathbf{s}%
_{N} \} $, $\mathbf{s}_{i}\in\mathcal{D}$, $N\geq1$.

In the next paragraph, the way the potential and the conditional
potential are modeled and estimated
is discussed.

\subsection{Model definition}\label{sec32}
The geostatistical potential model (GPM) is introduced here as
the main statistical tool for the analysis of spatial data arising from
concurrent measurements in the presence of interaction between the measuring
instruments. In its general form, the GPM is described by the following
hierarchy of equations:%
%
\begin{eqnarray}
\label{eqGPM} y(\mathbf{s};\mathcal{S}) & = & h_{\bolds{\vartheta}} \bigl( u (
\mathbf{s}%
),\mathbf{s},\mathcal{S} \bigr) ,
\nonumber
\\
u ( \mathbf{s} ) & = & q(\mathbf{s})+\varepsilon(\mathbf {s}) ,
\\
q(\mathbf{s}) & = & \mu+\mathbf{x}(\mathbf{s})\bolds{\beta}+\gamma w( 
\mathbf{s}).
\nonumber
\end{eqnarray}
At the first stage of (\ref{eqGPM}), $y(\mathbf{s};\mathcal{S})$ is
the measured conditional potential
at the spatial location $\mathbf{s}$ while $h_{\bolds{\vartheta
}}\dvtx\mathbb{R}\times\mathcal{D}\times\mathbb{S}\longrightarrow
\mathbb{R}$
is the \textit{interaction function} which is parametrized by the
parameter vector $\bolds{\vartheta}$.
The set $\mathbb{S}$ is the set of all finite spatial
point patterns over $\mathcal{D}$ including the nonsimple patterns
(i.e., patterns with overlapping points).
At the second stage, $\varepsilon(\mathbf{s})$ represents an
error component which is assumed to be i.i.d. $N(0,\sigma_{\varepsilon
}^{2})$ and is supposed to capture both the measuring error and the model
error. Finally, at the third stage, the potential $q(\mathbf{s})$ is
modeled by three summands, where $\mu$ is the mean, $\mathbf
{x}(\mathbf{s})$
is a vector of covariates, $\bolds{\beta}$ is the vector of coefficient,
$%
w(\mathbf{s})$ is a zero-mean latent Gaussian process and $\gamma$ is a
scale parameter. The covariance function of $w(\mathbf{s})$ is
$\operatorname{cov} ( w(%
\mathbf{s}),w(\mathbf{s}^{\prime}) ) =\rho_{\bolds{\theta
}} (
\mathbf{s},\mathbf{s}^{\prime} ) $, with $\rho_{\bolds{\theta}%
} ( \mathbf{s},\mathbf{s}^{\prime} ) $ a valid correlation
function parametrized by the vector $\bolds{\theta}$. The model parameter
vector is $\Psi= ( \mu,\bolds{\beta}^{\prime},\sigma_{\varepsilon
}^{2},\gamma,\bolds{\theta}^{\prime},\bolds{\vartheta}^{\prime}
) $
and it completely characterizes the GPM.

Note that, for the reasons discussed later on in the paper, it is
assumed that, conditionally to the observed covariates $\mathbf {x}
(\mathbf{s} )$, the observed $y(\mathbf{s};\mathcal{S})$ is not
preferentially sampled [see \citet{Diggle2010}] with respect to
the latent variable $w$. As a consequence, the set of spatial locations
$\mathcal{S}$ is treated as a constant rather than as the realization
of a spatial point process, the local density of which is driven by~$w$.

In order to have a better insight into the role of the interaction
function $%
h_{\bolds{\vartheta}}$, the following family of interaction functions is adopted:
%
\begin{equation}\label{eqinteractionfunction}
h_{\bolds{\vartheta}} \bigl( u ( \mathbf{s} ),\mathbf {s},\mathcal{S} \bigr) =u (
\mathbf{s} ) \cdot \biggl( 1+\sum_{\mathbf
{s}^{\prime
}\in\mathcal{S}}f_{\bolds{\vartheta}}
\bigl( \mathbf{s},\mathbf {s}^{\prime
} \bigr) \biggr)^{-1}=u (
\mathbf{s} ) \cdot g_{\bolds
{\vartheta}}%
(\mathbf{s};\mathcal{S}),
\end{equation}
where $f_{\bolds{\vartheta}} ( \mathbf{s},\mathbf{s}^{\prime
} )\dvtx\mathbb{R}^{2}\times
\mathbb{R}^{2}\longrightarrow
\mathbb{R}^{+}$ is a generic nonnegative binary function.

The function $f_{\bolds{\vartheta}} ( \mathbf{s},\mathbf
{s}^{\prime
} ) $ can be any continuous function but, for practical applications,
it should be monotonically decreasing with respect to distance. For instance,
%
\begin{equation}\label{eqffunction}
f_{\bolds{\vartheta}} \bigl( \mathbf{s},\mathbf{s}^{\prime} \bigr)
=f_{%
\bolds{\vartheta}} \bigl( \bigl\llVert \mathbf{s}-\mathbf{s}^{\prime
}\bigr
\rrVert \bigr) =\exp \biggl( -\frac{\llVert \mathbf
{s}-\mathbf{s}%
^{\prime}\rrVert }{\phi} \biggr)^{\alpha},
\end{equation}
where \mbox{$\llVert \cdot\rrVert $} is the Euclidean
distance and $\bolds{\vartheta}=
( \phi,\alpha )^{\prime}$ is the function parameter
vector. In equation (\ref{eqffunction}), $\phi$ defines the strength of the interaction while
$%
\alpha>0$ is a shape parameter.\vadjust{\goodbreak}

Note that%
%
\begin{eqnarray}
\label{eqGPM2} y(\mathbf{s};\mathcal{S}) & = & u ( \mathbf{s} ) \cdot
g_{\bolds{\vartheta}}(\mathbf{s};\mathcal{S})
\nonumber
\\
& = & q(\mathbf{s})\cdot g_{\bolds{\vartheta}}(\mathbf{s};\mathcal {S})+\varepsilon(
\mathbf{s})\cdot g_{\bolds{\vartheta}}(\mathbf {s};\mathcal{S})
\\
& = & q(\mathbf{s};\mathcal{S})+\varepsilon(\mathbf{s};\mathcal{S}),
\nonumber
\end{eqnarray}
namely, the observed potential is equal to the conditional
potential $q(\mathbf{s};\mathcal{S})$ plus a transformation of the
error $\varepsilon(\mathbf{s})$.
In particular, the second line of equation (\ref{eqGPM2}) follows
directly from the second
stage of model (\ref{eqGPM}), while the third line is the second line
rewritten in a more compact notation.

The term $g_{\bolds{\vartheta}}(\mathbf{s};\mathcal{S})$ is the key
element of the interaction function and it deserves
more explanation. If, as an example, the function (\ref{eqffunction}) is
considered and \mbox{$\mathcal{S}\equiv\varnothing$}, namely, if there are
no measuring instruments,
then $g_{\bolds{\vartheta}}(\mathbf{s};\mathcal{S})=1$ since the
summand in equation (\ref{eqinteractionfunction}) cannot be evaluated and it is equal to zero by
definition. When a measuring instrument is added, $\mathcal{S}= \{
\mathbf{s}_{1} \} $, the potential at $\mathbf{s}$ is a
function of the distance between $\mathbf{s}$ and $\mathbf{s}_{1}$. In
particular, if $\mathbf{s=s}_{1}$, then $g_{\bolds{\vartheta}}(\mathbf
{s};\mathcal{S})=0.5$.
On the contrary, when $\llVert \mathbf{s}-\mathbf{s}_{1}\rrVert
\rightarrow\infty$, then $g_{\bolds{\vartheta}}(\mathbf{s};\mathcal
{S})\rightarrow1$. This
reflects the fact that the action of absorbing the potential at site $%
\mathbf{s}_{1}$ influences the measure at the site~$\mathbf{s}$. It
is worth
noting that $\mathbf{s}$ and $\mathbf{s}_{1}$ are exchangeable in the sense
that absorbing and measuring the potential are equivalent actions and that
the potential cannot be measured without being absorbed.

In this work, the measuring instruments are supposed to be
\textit{equally-effective}, that is,
$g ( \mathbf{s}_{i}; \{ \mathbf{s}%
_{j} \}  ) =g ( \mathbf{s}_{j}; \{ \mathbf
{s}_{i} \}
)$ for all $\llVert \mathbf{s}_{i}-\mathbf{s}_{j}\rrVert $.
The property of equally-effectiveness is satisfied if the binary
function $f_{\bolds{\vartheta}%
} ( \mathbf{s},\mathbf{s}^{\prime} ) $ is
commutative,\footnote{The binary function $f$ is commutative if
$f(x,y)=f(y,x)$.} which is
the case of the function (\ref{eqffunction}). In practical
applications, the property
may not be satisfied in the sense that a measuring instrument might be
more effective
in absorbing the potential than a second instrument close to it.
Suppose equally-effective measuring instruments, however, simplify the
model and any
discrepancy from it is accounted for by the error term $\varepsilon$.
Note that the measure of effectiveness is strictly related to the
measure of attractiveness of
the \textit{spatial behavior of consumers models} typical of the
geomarketing literature [see \citet{cliquet2006}].

\begin{figure}

\includegraphics{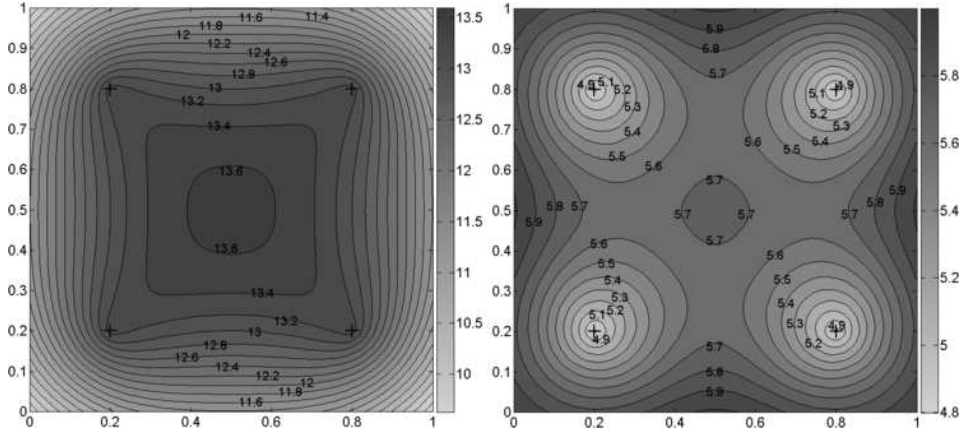}

\caption{(Left) potential $q(\mathbf{s})$; (right) conditional
potential $q(\mathbf{s};\mathcal{S})$.}
\label{potentialpattern}
\end{figure}

To better understand the notions of potential and conditional
potential, the
following example is considered. Suppose that four
measuring instruments are located at $%
\mathbf{s}_{1}=(0.2,0.2)$, $\mathbf{s}_{2}=(0.2,0.8)$, $\mathbf{s}%
_{3}=(0.8,0.2)$ and $\mathbf{s}_{4}=(0.8,0.8)$, $\mathbf{s}_{i}\in
\mathcal{D%
}\equiv\lbrack0,1]\times\lbrack0,1]$, $i=1,\ldots,4$. The GPM
considered is%
%
\begin{eqnarray}
\label{eqGIMexample} y(\mathbf{s};\mathcal{S}) & = & u(\mathbf{s})\cdot
g_{\bolds{\vartheta
}}(\mathbf{%
s};\mathcal{S})=q(\mathbf{s};\mathcal{S}) ,
\nonumber
\\
u(\mathbf{s}) & = & q(\mathbf{s}) ,
\\
q(\mathbf{s}) & = & w(\mathbf{s}),
\nonumber
\end{eqnarray}
namely, it is supposed that the conditional potential $q(\mathbf
{s};\mathcal{S%
})$ is observed without error. Furthermore, suppose that%
%
\begin{eqnarray}
\label{eqrhoexample}
\rho_{\bolds{\theta}} \bigl( \mathbf{s},\mathbf{s}^{\prime} \bigr) &=&
\rho_{\theta} \bigl( \bigl\llVert \mathbf{s}-\mathbf{s}^{\prime
}\bigr
\rrVert \bigr) =\exp \biggl( -\frac{\llVert \mathbf
{s}-\mathbf{s}%
^{\prime}\rrVert }{0.8} \biggr) ,
\\
\label{eqfexample}
f_{\bolds{\vartheta}} \bigl( \mathbf{s},\mathbf{s}^{\prime} \bigr)
&=&f_{\bolds{\vartheta} } \bigl( \bigl\llVert \mathbf{s}-\mathbf {s}^{\prime
}\bigr
\rrVert \bigr) =\exp \biggl( -\frac{\llVert \mathbf
{s}-\mathbf{s}%
^{\prime}\rrVert }{0.3} \biggr)
\end{eqnarray}
and that $y(\mathbf{s}_{i};\mathcal{S}\setminus\mathbf
{s}_{i})=10$.\footnote{Note that, in general, $y(\mathbf{s};\mathcal
{S})$ should be simulated following equation (\ref{eqsim}). In this
case, in order to better appreciate the role of the interaction
function, $y(\mathbf{s}_{i};\mathcal{S}\setminus\mathbf{s}_{i})$ is
supposed to be equal for all the locations.} The estimated potential and
conditional potential are reported in the left and in the right parts
of Figure
\ref{potentialpattern}, respectively. Regarding the potential,
its value at the measuring instrument locations is equal to $13.2>10$. Each
%
measuring instrument measures a potential equal to $10$ since a
fraction of
it is absorbed by the remaining measuring instruments.
Indeed, the potential $q(\mathbf{s}_{i})=13.2$ would be measured by
the single
measuring instrument if the other instruments were not present. The conditional
potential, as expected, has its lowest value at the measuring
instrument locations
and represents the potential that would be observed by a fifth
measuring instrument
if placed at the generic $\mathbf{s}$.

\section{Parameter estimation and inference}
\label{secestimationandinference}\label{sec4}

Let $\mathbf{y}\equiv\mathbf{y} ( \mathcal{S} ) $ be the
$N\times
1 $ vector of data collected at the sampling sites $\mathcal{S}$. The
measurement equation for the vector $\mathbf{y}$ is%
%
\begin{equation}\label{eqmeasurementequationonsite}
\mathbf{y}=\mathbf{G} ( \bolds{1}\mu+\mathbf{X}\bolds{\beta}+\gamma
\mathbf{w}+\bolds{\varepsilon} ),
\end{equation}
where $\bolds{1}$ is the $N\times1$ vector of ones, $\mathbf
{X}\equiv
\mathbf{X} ( \mathcal{S} ) $ is the $N\times b$ matrix of
covariates, $\mathbf{w}\equiv\mathbf{w} ( \mathcal{S} )
$ is the
latent Gaussian process at $\mathcal{S}$ with variance--covariance
matrix $%
\bolds{\Sigma}_{\mathbf{w}}\equiv\bolds{\Sigma}_{\mathbf{w}} (
\mathcal{S},\bolds{\theta} ) $ and $\bolds{\varepsilon}\equiv
\bolds{\varepsilon} ( \mathcal{S} ) $ is the measurement
error at
$\mathcal{S}$ with diagonal variance--covariance matrix $\bolds{\Sigma}_{
\bolds{\varepsilon}}=\sigma_{\varepsilon}^{2}I_{N}$. Finally, $\mathbf
{G}%
\equiv\mathbf{G}_{\bolds{\vartheta}} ( \mathcal{S} ) $ is
the $N\times
N$ diagonal matrix whose diagonal vector is
\[
\mathbf{g}= \bigl( g_{\bolds{\vartheta}}(\mathbf{s}_{1};\mathcal {S}
\setminus%
\mathbf{s}_{1}),\ldots,g_{\bolds{\vartheta}}(
\mathbf{s}_{N};\mathcal{%
S}\setminus\mathbf{s}_{N})
\bigr).
\]

Furthermore, suppose that $\mathcal{S}$ is partitioned as $ \{
\mathcal{S%
}^{(1)},\mathcal{S}^{(2)} \} $, where $\mathcal{S}^{(1)}$ is
the set of
sites where the data are available and $\mathcal{S}^{(2)}$ is the set of
sites where the data are missing. According to this, the vector
$\mathbf{y}$
is partitioned as $\mathbf{y}^{\ast}= ( \mathbf
{y}^{(1)},\mathbf{y}^{(2)} )^{\prime}$,
where $\mathbf{y}^{(1)}=\mathbf{Ly}$ is the subvector of the nonmissing
data and $\mathbf{L}$ is the appropriate elimination matrix. The
vector $%
\mathbf{y}^{\ast}$ is a permutation of $\mathbf{y}$ and $\mathbf
{y}=\mathbf{%
D}\mathbf{y}^{\ast}$, with $\mathbf{D}$ the proper commutation
matrix. The partitioned measurement equation becomes
\[
\mathbf{y}^{(i)}=\mathbf{G}^{(i)} \bigl( \bolds{1}^{(i)}
\mu +\mathbf{X}%
^{(i)}\bolds{\beta}+\gamma\mathbf{w}^{(i)}+
\bolds{\varepsilon}%
^{(i)} \bigr);\qquad i=1,2,
\]
and the variance--covariance matrix of the permuted errors is conformably
partitioned as%
\[
\operatorname{Var} \bigl[ \bigl( \bolds{\varepsilon}^{(1)},\bolds{\varepsilon}%
^{(2)}
\bigr)^{\prime} \bigr] =\pmatrix{ \mathbf{R}_{11} &
\mathbf{R}_{12}
\cr
\mathbf{R}_{21} & \mathbf{R}_{22}}.
\]
In the sequel, given $\mathbf{b}$ a generic vector and $\mathbf{B}$ a
generic matrix, $\mathbf{b}^{(1)}$ and $\mathbf{B}^{(1)}$ will stand
for $%
\mathbf{Lb}$ and $\mathbf{LBL}^{\prime}$, respectively, bearing
in mind
that, in general, $\mathbf{LB}^{-1}\mathbf{L}^{\prime}\neq (
\mathbf{LBL%
}^{\prime} )^{-1}$.

Given the data vector $\mathbf{y}$ and considering the GPM, the
following inferential problems are of interest:

\begin{longlist}[(5)]
\item[(1)] to provide an estimate of the model parameter vector $\Psi$;

\item[(2)] to provide confidence intervals for the elements of $\hat{\Psi}$;

\item[(3)] to estimate the potential $q ( \mathbf{s} ) $ over
the region $%
\mathcal{D}$ and its uncertainty;

\item[(4)] to estimate the conditional potential $q(\mathbf{s};\mathcal
{S})$ over
the region $\mathcal{D}$ and its uncertainty;

\item[(5)] to evaluate the expected total potential measured by a maximum of
measuring instruments.
\end{longlist}

\subsection{Parameter estimation}\label{parmodelestimation}\label{sec41}

Problem 1 is tackled here following the maximum likelihood (ML) approach.
With $w ( \mathbf{s} ) $ being a latent process and due to possible
missing data, the expectation--maximization (EM) algorithm is adopted to find
the ML estimate $\hat{\Psi}$ of $\Psi$.

The EM algorithm is based on the complete-data likelihood function
$L_{\Psi
} ( \mathbf{y},\mathbf{w} ) $ and it provides an iterative
procedure to update the model parameter estimate from $\hat{\Psi
}^{(k)}$ to $%
\hat{\Psi}^{(k+1)}$ until convergence [see \citet{McLachlan2008}].
In particular, for each iteration of\vadjust{\goodbreak}
the algorithm, the E-step computes the conditional expectation%
\[
Q \bigl( \Psi,\hat{\Psi}^{(k)} \bigr) =E_{\hat{\Psi}^{(k)}} \bigl[
L_{\Psi
} ( \mathbf{y},\mathbf{w} ) \mid\mathbf{y}^{(1)} \bigr],
\]
while, at the M-step, the following maximization is carried out:%
\[
\hat{\Psi}^{(k+1)}=\mathop{\arg\max}_{\Psi}Q \bigl( \Psi,\hat {
\Psi}^{(k)} \bigr),
\]
which is equivalent to solve the equation%
%
\begin{equation}\label{eqmaximizationstep}
\frac{\partial Q ( \Psi,\hat{\Psi}^{(k)} ) }{\partial
\Psi}=%
\mathbf{0}.
\end{equation}

Considering the approach described in \citet{Fassofinazzi2011},
the following closed form updating formulas have been derived:
%
\begin{eqnarray}
\label{equpdatesigmaeps}
\hat{\mu}^{(k+1)} &=&\frac{\operatorname{tr} [  ( \hat
{\mathbf{e}}^{(1)}+\mu^{(k)}\bolds{1}^{(1)} )  ( \bolds{1}^{(1)} ){}^{\prime} ] }{N-N_{m}} ,
\\
\hat{\bolds\beta}^{(k+1)} &=& \bigl[ \bigl( \mathbf{X}^{(1)}
\bigr)^{\prime}\mathbf{X}^{(1)} \bigr]^{-1} \bigl( \mathbf
{X}^{(1)} \bigr)^{\prime}\cdot \bigl( \hat{\mathbf{e}}^{(1)}+
\mathbf{X}^{(1)}\bolds{\beta}%
^{(k)} \bigr) ,
\\
\label{equpdatebeta}
\bigl( \hat{\sigma}_{\varepsilon}^{2} \bigr)^{(k+1)} &=&
\frac{1}{N}\operatorname{tr} \pmatrix{ \hat{\mathbf{e}}^{(1)}\cdot \bigl(
\hat{\mathbf{e}}^{(1)} \bigr)^{\prime
}+ \bigl( \gamma^{(k)}
\bigr){}^{2}\hat{\mathbf{A}}^{(1)} & \mathbf{0}
\vspace*{1pt}\cr
\mathbf{0} &
\mathbf{R}_{22}},
\\
\label{equpdategamma}
\hat{\gamma}{}^{(k+1)} &=&\frac{\operatorname{tr} [  ( \hat{\mathbf{e}}%
^{(1)}+\gamma^{(k)}\hat{\mathbf{w}}^{(1)} )  (
\hat{\mathbf{w}}%
^{(1)} )^{\prime} ] }{\operatorname{tr} [ \hat{\mathbf{w}}%
^{(1)} ( \hat{\mathbf{w}}^{(1)} )^{\prime}+\hat
{\mathbf{A}}^{(1)}%
] },
\end{eqnarray}
where $\hat{\mathbf{e}}^{(1)}=  ( \mathbf{G}^{(1)} )^{-1}
\mathbf{y}^{(1)}\mathbf{-}\mu^{(k)}\bolds{1}^{(1)}-\mathbf
{X}^{(1)}%
\bolds{\beta}^{(k)}-\gamma^{(k)}\hat{\mathbf{w}}^{(1)}$, $N_{m}$
is the number of missing data in $\mathbf{y}$ and
%
\begin{eqnarray}
\label{eqestimatedlatent}
\hat{\mathbf{w}} &=&E_{\Psi^{(k)}} \bigl( \mathbf{w}\mid\mathbf{y}
^{(1)} \bigr),
\\
\label{eqestimatedlatentvariance}
\hat{\mathbf{A}} &=&\operatorname{Var}_{\Psi^{(k)}} \bigl( \mathbf{w}\mid\mathbf
{y}%
^{(1)} \bigr)
\end{eqnarray}
are the estimated latent variable and the estimation variance,
respectively. The evaluation of (\ref{eqestimatedlatent}) and
(\ref{eqestimatedlatentvariance}) is reported in Appendix~\ref{appA}.

The remaining model parameters can be updated by numerical optimization solving
$ ( \hat{\bolds{\theta}}^{(k+1)},\hat{\bolds{\vartheta
}}^{(k+1)} ) =
\mathop{\arg\max}_{\bolds{\theta},\bolds{\vartheta}}Q ( \Psi,
\hat{\Psi}^{(k)} )$.
If both the correlation function $\rho_{\bolds{\theta}}$ and
the interaction function $h_{\bolds{\vartheta}}$ have analytical form of the
first and second derivative with respect to $\bolds{\theta}$ and
$\bolds{\vartheta}$,
respectively, both the parameters can be updated by adapting the
algorithm given in \citet{XuWikle2007}.\looseness=-1

Before concluding the paragraph, a point that is worth mentioning is
how the preferential sampling problem
can affect the model parameter estimation in the case of the GPM.
The spatial data $\mathbf{y} ( \mathcal{S} )$ are
preferentially sampled with respect
to the potential $q(\mathbf{s})$ if the spatial pattern of $\mathcal
{S}$ is not independent of
$q(\mathbf{s})$. In practice, the spatial density of $\mathcal{S}$
can be higher at the spatial
locations where $q(\mathbf{s})$ is known or expected to be high.

As discussed in \citet{Diggle2010}, if the data are preferentially
sampled and the issue is not\vadjust{\goodbreak}
addressed, then the estimation of the parameter $\bolds{\theta}$ related
to the latent
variable $w(\mathbf{s})$ is generally biased. With respect to the GPM,
however, the model
considered in \citet{Diggle2010} does not include covariates. If the
data are preferentially
sampled with respect to the potential $q(\mathbf{s})$ but the
covariates explain a good part
of the variability of the potential, then $w(\mathbf{s})$ models only
the ``residual'' random
field $\tilde{e}(\mathbf{s})=q(\mathbf{s})-\mathbf{x}(\mathbf
{s})\hat{\bolds\beta}$ and the data
$\mathbf{y} ( \mathcal{S} )-\mathbf{X} ( \mathcal
{S} )\hat{\bolds\beta}$ can be
assumed to be not preferentially sampled with respect to~$\tilde
{e}(\mathbf{s})$.
In other words, even when the data are preferentially sampled with
respect to $q(\mathbf{s})$,
the adoption of good (spatial) covariates largely mitigates the problem.
If no covariates are available and the data are suspected to be
preferentially sampled, then
the approach in \citet{Diggle2010} should be considered.

\subsection{Parameter confidence intervals}\label{parconfidenceintervals}\label{sec42}

As known, the classic EM algorithm does not provide information about
the uncertainty of the estimated parameter vector $\hat{\Psi}$. In
order to avoid the more cumbersome supplemented EM algorithm [see
\citet{Mengrubin1991}], two methods are proposed to solve problem
2 of the above list, namely, to provide confidence intervals for the
elements of $\hat{\Psi}$.

The first method is based on the fact that the maximum likelihood estimator
has asymptotically normal distribution $N ( \Psi_{0},\mathbf{I}%
^{-1} ) $, with $\Psi_{0}$ the ``true'' value of $\Psi$ and
$\mathbf{I}$
the Fisher information matrix. An approximation of the information matrix
for multivariate normal variables can be evaluated as%
%
\begin{eqnarray}\label{eqinformationmatrix}
\tilde{\mathbf{I}}_{ij} &=&\partial_{i}\bolds{
\epsilon}^{\prime}
\bolds{\Sigma}_{\bolds{\epsilon}}^{-1}
\partial_{j}\bolds{\epsilon}+\tfrac{1}{2} 
\operatorname{tr} \bigl( \bolds{
\Sigma}_{\bolds{\epsilon}}^{-1}\partial_{i}%
\bolds{
\Sigma}_{\bolds{\epsilon}}\bolds{\Sigma}_{\bolds{\epsilon}%
}^{-1}
\partial_{j}\bolds{\Sigma}_{\bolds{\epsilon}} \bigr)
\nonumber\\[-8pt]\\[-8pt]
&&{} +\tfrac{1}{4}\operatorname{tr} \bigl( \bolds{\Sigma}_{\bolds{\epsilon}%
}^{-1}
\partial_{i}\bolds{\Sigma}_{\bolds{\epsilon}} \bigr) \operatorname{tr}%
\bigl( \bolds{
\Sigma}_{\bolds{\epsilon}}^{-1}\partial_{j}\bolds{\Sigma
}_{\bolds{\epsilon}} \bigr)
\nonumber
\end{eqnarray}
[see \citet{Shumwaystoffer2006}], where $\partial_{i}\bolds{\epsilon}$
and $\partial_{i}\bolds{\Sigma}_{\bolds{\epsilon}}$ are short notation
for $\partial\bolds{\epsilon} ( \Psi ) /\partial\Psi_{i}$ and
$\partial\bolds{\Sigma}_{\bolds{\epsilon}} ( \Psi )
/\partial
\Psi_{i}$, respectively, and $1\leq i,j\leq\llvert \Psi\rrvert $.

In the case of the GPM, the vector
%
\begin{equation}\label{eqinnovation}
\bolds{\epsilon}=\mathbf{y-G} ( \bolds{1}\mu+\mathbf{X }\bolds{\beta}%
)
\end{equation}
is normally distributed with variance--covariance matrix%
%
\begin{eqnarray}\label{eqinnovationvariance}
\bolds{\Sigma}_{\bolds{\epsilon}} &=&\operatorname{Var} \bigl( \mathbf{y-G} ( \bolds{1}\mu+
\mathbf{X }\bolds{\beta} ) \bigr)
\nonumber\\
&=&\operatorname{Var} ( \gamma\mathbf{Gw+G}\bolds{\varepsilon} )
\nonumber\\[-8pt]\\[-8pt]
&=&\mathbf{G} \bigl( \gamma^{2}\bolds{\Sigma}_{\mathbf{w}}+
\bolds{\Sigma}%
_{\bolds{\varepsilon}} \bigr) \mathbf{G}^{\prime}
\nonumber
\\
&=&\mathbf{gg}^{\prime}\odot \bigl( \gamma^{2}\bolds{\Sigma
}_{\mathbf{w}%
}+\bolds{\Sigma}_{\bolds{\varepsilon}} \bigr),
\nonumber
\end{eqnarray}
where $\odot$ is the Hadamard product operator. The solution for the
derivatives $\partial_{i}\bolds{\epsilon}$ and $\partial_{i}
\bolds{\Sigma}_{\bolds{\epsilon}}$ is reported in Appendix~\ref{appB}. In the
presence of
missing data, equation (\ref{eqinformationmatrix}) is still valid,
but $\bolds{\epsilon}$ and $\bolds{\Sigma}_{\bolds{\epsilon}}$ have to be replaced
with $\bolds{\epsilon}^{(1)}$ and $\bolds{\Sigma}_{\bolds{\epsilon}%
}^{(1)}$, respectively.

With $\tilde{\mathbf{I}}$ available, approximated confidence
intervals for
the elements of $\hat{\Psi}$ are immediately provided by considering $
N ( \hat{\Psi},\tilde{\mathbf{I}}^{-1} ) $. Note,
however, that\vadjust{\goodbreak} $%
N ( \hat{\Psi},\tilde{\mathbf{I}}^{-1} ) $ is a good
approximation
of the distribution $ [ \Psi\mid\mathbf{y} ( \mathcal
{S} ) %
] $ only when $N$ is large, which may not be the case in practical
applications.

To solve this problem, following \citet{Fassocameletti2010}, a
second method based on the bootstrap technique is considered. Let
$\hat{\Psi}$ be the estimated parameter vector. For each
bootstrap run $%
m$, the vector $\mathbf{y}_{(m)}=\mathbf{D}
\bigl[{\mathbf{y}_{(m)}^{(1)}}\enskip{\mathbf{y}^{(2)}}\bigr]^{\prime}$ is considered, where
%
\begin{equation}\label{eqsim}
\mathbf{y}_{(m)}^{(1)}=\mathbf{L}\cdot\mathbf{G}\cdot (
\bolds{1}\hat{\mu}+\mathbf{X}\hat{\bolds\beta}%
+\hat{\gamma}
\tilde{\mathbf{w}}_{(m)}+\tilde{\bolds\varepsilon}_{(m)} )
\end{equation}
and where $\tilde{\mathbf{w}}_{(m)}$ and $\tilde
{\bolds\varepsilon}_{(m)}$ are
realizations from the multivariate normal distributions $N (
0,\bolds{\Sigma}_{\bolds{\varepsilon}} ( \hat{\sigma}_{\varepsilon
}^{2} )  ) $ and $N ( 0,\bolds{\Sigma}_{\mathbf
{w}} (
\hat{\bolds{\theta}} )  )$, respectively. Note that
$\mathbf{y}%
_{(m)}$ preserves the missing data pattern of the observed $\mathbf
{y}$. The
simulated $\mathbf{y}_{(m)}$ is used to estimate a new parameter
vector $%
\hat{\Psi}_{(m)}$ through the EM algorithm and the set%
%
\begin{equation}\label{eqparametersample}
\hat{\Psi}_{s}= \{ \hat{\Psi}_{(1)},\ldots,\hat{\Psi
}_{(M)} \}
\end{equation}
is considered as a sample from the distribution $ [ \Psi\mid
\mathbf{y}%
( \mathcal{S} )  ] $. If $M$ is large enough, then
$\hat{\Psi}%
_{s}$ can be used to derive approximated confidence intervals for the
elements of $\hat{\Psi}$ without normality assumptions.

\subsection{Potential and conditional potential estimation}
\label{parunconditionalpotential}\label{sec43}

Following the plug-in approach, the estimated
potential is obtained as%
%
\begin{equation}\label{eqkriging}
q_{\hat{\Psi}}(\mathbf{s})=\hat{\mu}+\mathbf{x} ( \mathbf {s} )
\hat{\bolds\beta}+\hat{\gamma}\hat{\mathbf{w}} ( \mathbf {s} );\qquad
\mathbf{s}\in
\mathcal{D},
\end{equation}
where $\hat{\mathbf{w}} ( \mathbf{s} ) =E_{\hat{\Psi
}} (
\mathbf{w}(\mathbf{s})\mid\mathbf{y} ) $ is the kriging
estimate of $%
\mathbf{w} ( \mathbf{s} ) $, which is evaluated
analogously to $%
\hat{\mathbf{w}}$ in equation (\ref{eqestimatedlatent}).

The uncertainty of $q_{\hat{\Psi}}(\mathbf{s})$ is directly related
to the
uncertainty of $\hat{\Psi}$ which is expressed by $ [ \Psi\mid
\mathbf{y%
} ( \mathcal{S} )  ] $. Again, approximated confidence
intervals on $q_{\hat{\Psi}}(\mathbf{s})$ can be provided by repeatedly
estimating $q_{\Psi}(\mathbf{s})$ with $\Psi$ extracted either from $
N ( \hat{\Psi},\tilde{\mathbf{I}}^{-1} ) $ or from the
set $\hat{%
\Psi}_{s}$ defined in equation (\ref{eqparametersample}). The estimated
conditional potential is simply given by
\[
q_{\hat{\Psi}}(\mathbf{s};\mathcal{S})=q_{\hat{\Psi}}(\mathbf {s})\cdot
g_{%
\hat{\bolds{\vartheta}}}(\mathbf{s};\mathcal{S});\qquad \mathbf{s}\in \mathcal{D}.
\]
Approximated confidence intervals on $q_{\hat{\Psi}}(\mathbf
{s};\mathcal{S})$
are provided following the same approach for $q_{\hat{\Psi}}(\mathbf{s})$.

\subsection{Total potential estimation}\label{partotalpotential}\label{sec44}

The conditional potential $q(\mathbf{s};\mathcal{S})$ provides
information about
the expected observation when a measuring instrument is placed at the generic
location $\mathbf{s}$ given the existence of the other instruments. In
practice,
the following quantity is also of interest:
%
\begin{equation}\label{eqoptnet}
v=\max_{\mathcal{S} \in\mathbb{S}}\sum_{\mathbf
{s}^{\prime
}\in\mathcal{S}}q \bigl(
\mathbf{s}^{\prime};\mathcal{S} \setminus \mathbf{s}^{\prime}
\bigr);\qquad
\mathcal{S}\neq\varnothing.
\end{equation}

If, for example, $q(\mathbf{s})$ is the spatial market potential, then
$v$ represents the
maximum market volume for the trading area $\mathcal{D}$. Note that
$v$ cannot be
obtained by simply integrating $q(\mathbf{s})$ or $q(\mathbf
{s};\mathcal{S})$ over $\mathcal{D}$.

A simple way to estimate $v$ is to consider the estimated conditional
potential $q_{\hat{\Psi}}(\mathbf{s};\mathcal{S})$
with $\mathcal{S}=\varnothing$ and to sequentially populate $\mathcal
{S}$ by choosing the spatial
location of $\mathcal{D}$ where $q_{\hat{\Psi}}(\mathbf{s};\mathcal
{S})$ is maximum for the current $\mathcal{S}$.
Following this approach, an estimation of $v$ is obtained for $\llvert \mathcal{S}\rrvert \rightarrow\infty$. In practice,
the value of $v$ stops to increase significantly after a finite number
of iterations.
Note that a by-product of (\ref{eqoptnet}) is the optimum $\mathcal
{S}$ with respect
to the maximization of $v$. When the main aim is the optimization of a
retail network, however,
the above approach should be adapted in order to impose a threshold on
the minimum (geographic) distance between two
elements of~$\mathcal{S}$.

\section{Case study}\label{seccasestudy}\label{sec5}

The GPM is considered here in order to estimate the market potential of
an economic daily
newspaper over the area of the city of Bergamo, northern Italy. The aim
is to identify
the areas with the highest market potential, to identify the areas
where it would be profitable
to open additional newsstands and to evaluate the maximum total market
volume for the
city.

The Italian daily newspaper market is characterized by $64$ main
newspaper heads with
an average market volume of around $5.5$ million daily copies. As far
as the city of Bergamo
concerns, only $16$ out of $64$ newspaper heads are commonly
commercialized, as most of them
are local heads referring to other Italian cities.
The economic newspaper considered in this study represents $8\%$ of the
total sales volume
for the Bergamo area in terms of daily copies. Moreover, it should be
noted that
the economic newspaper is of a clientele which differs from that of the
most popular newspapers.
This implies that the market potential of the economic newspaper is not
necessarily reflected
in the spatial distribution of the newsstands. In other words,
conditionally to the observed covariates, the sales data are not preferentially
sampled with respect to the market potential of the newspaper. This is
also justified by the
fact that the sale of daily newspapers represent only $20\%$ of the
total revenue of a newsstand.

\begin{figure}

\includegraphics{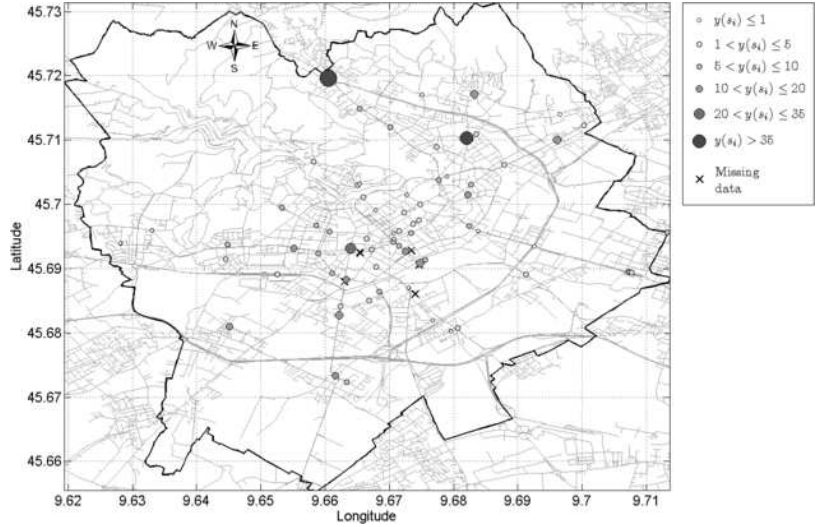}

\caption{Newsstand locations and circle plot of the working day
average daily number of copies sold.}
\label{newsstandlocations}
\end{figure}

The data available for the study consist of the yearly average daily
number of copies sold
on working days by $N=75$ newsstands located over the Bergamo area. The
sales data of $5$
newsstands are unavailable though their location is known. The total
daily average sales volume for
the available newsstands is around $491$ copies\vadjust{\goodbreak} and it is believed that
the maximum
total volume attainable for the city of Bergamo is higher.
The newsstand spatial locations are shown in Figure~\ref{newsstandlocations}, along with the
circle-plot of the average daily number of copies sold.

By considering the interpretation introduced in Section~\ref{parnewapproach},
it can be stated that the measuring system is represented by the
newsstands and that
the interaction between the measuring instruments is of the absorption
type. In fact, once the customer has bought a copy of the
newspaper, it is absorbed in the sense that the same customer will not buy
(during the same day) the same copy of the newspaper, neither at the same
nor at a different newsstand.
Since the newspaper price is fixed, the newsstands are considered
equally effective and
it is supposed that the customer chooses the nearest newsstand.
Customer loyalty is admitted,
but it is supposed that a newsstand is not more attractive than another.

The GPM model considered is the same defined in (\ref{eqGPM}) but
with $\mu\equiv0
$. This implies that the market potential goes to zero when moving far
from the newsstand network, as $w(\mathbf{s})$ converges
to its marginal mean which is zero. It follows that the market
potential is zero (or very close to zero) over the areas where
it would be unfeasible to have a newsstand.
The spatial correlation function of the latent component $w$ is chosen to
be
%
\begin{equation}\label{eqrhosim}
\rho_{\theta} \bigl( \mathbf{s},\mathbf{s}^{\prime} \bigr) =\exp
\biggl( -%
\frac{\llVert \mathbf{s}-\mathbf{s}^{\prime}\rrVert
}{\theta}%
\biggr),
\end{equation}
while the function (\ref{eqinteractionfunction}) is considered as
the interaction function, with
%
\begin{equation}\label{eqfsim}
f_{\phi} \bigl( \mathbf{s},\mathbf{s}^{\prime} \bigr) =\exp \biggl(
\frac{%
-\llVert \mathbf{s}-\mathbf{s}^{\prime}\rrVert }{\phi
} \biggr).
\end{equation}

\begin{figure}

\includegraphics{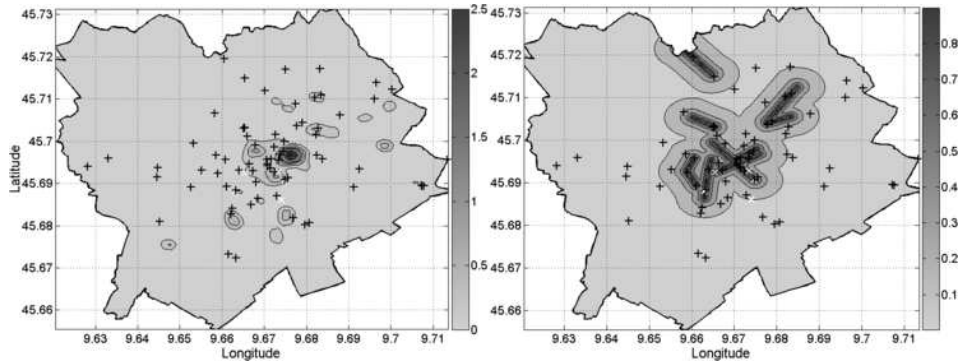}

\caption{Model covariates: (left) spatial density of joint-stock
companies; (right) function of the geographic distance to the nearest
busy street section.}
\label{covariates}
\end{figure}

Two covariates are considered. The first covariate represents the
spatial density
of the joint-stock companies with registered offices in Bergamo. The companies
are expected to induce a higher sales volume at the near newsstands.
The second covariate is a function of the minimum Euclidean distance
to the busiest street sections in terms of people and car traffic. In
particular, the
covariate value at the generic location $\mathbf{s}$ is given by
$1/(d_{\mathrm{min}}(\mathbf{s})+0.1)$, where
$d_{\mathrm{min}}(\mathbf{s})$ is the Euclidean distance (expressed in
kilometers) from the location
$\mathbf{s}$ to the nearest street section.
Both the covariates are depicted in Figure~\ref{covariates}. In order
to make the
$\bolds{\beta}$ coefficients directly comparable, the covariates at the
newsstand locations
have been rescaled to the range $[0,1]$.

\begin{table}[b]
\tablewidth=273pt
\caption{Estimated model parameter and $95\%$ bootstrap confidence
interval}\label{tabpsiestimated}%
\begin{tabular*}{\tablewidth}{@{\extracolsep{\fill}}ld{2.2}d{2.2}
d{3.2}d{2.2}d{3.2}d{3.2}@{}}
\hline
& \multicolumn{1}{c}{$\bolds{\hat{\beta}_{1}}$}
& \multicolumn{1}{c}{$\bolds{\hat{\beta}_{2}}$}
& \multicolumn{1}{c}{$\bolds{\hat{\sigma}_{\varepsilon}^{2}}$}
& \multicolumn{1}{c}{$\bolds{\hat{\gamma}}$}
& \multicolumn{1}{c}{$\bolds{\hat{\theta}}$}
& \multicolumn{1}{c@{}}{$\bolds{\hat{\phi}}$} \\
\hline
Estimated & 18.29 & 27.65 &
11.77 & 14.63 &
81.77 & 231.69 \\
LCL & -0.08 & 19.19 &
3.22 & 11.17 &
14.92 & 195.73 \\
UCL & 66.91 & 66.51 &
103.33 & 31.60 &
253.06 & 474.64\\
\hline
\end{tabular*}
\end{table}

The model parameter vector $\Psi$ is estimated by means of the EM
algorithm as
discussed in Section~\ref{parmodelestimation} and by using the
software provided in \citet{Finazzi2012}. The estimation result is reported
in Table~\ref{tabpsiestimated} with confidence intervals evaluated by
following the bootstrap approach discussed in Section~\ref{parconfidenceintervals}
and $M=922$. Namely, $95\%$ confidence intervals are obtained by evaluating
empirical distributions on $\hat{\Psi}_{s}= \{ \hat{\Psi
}_{(1)},\ldots,\hat{%
\Psi}_{(M)} \} $. Note that the original number of bootstrap
runs was $M=1000$,
but $78$ runs have been ignored after testing the estimated parameters
against anomalous values.
In fact, the EM algorithm is not guaranteed to converge to the global
maximum of
the likelihood function. In this particular application, the
condition
$\phi>1500m$ has
been considered to identify bad estimation results, as values of $\phi
$ higher than $1500m$ implies a very strong
and unrealistic competition between the newsstands.

The empirical variance--covariance matrix of $\hat{\Psi%
}$ is reported in Table~\ref{tabpsivarcov} and it can be compared
with the
approximated Hessian matrix evaluated by considering equation (\ref{eqinformationmatrix}) and
reported in Table~\ref{tabhessian}. In particular, it can be noted
that the
approximated Hessian matrix tends to underestimate the variances related
to the elements of $\hat{\Psi}$.

\begin{table}
\tablewidth=290pt
\caption{Empirical variance--covariance matrix of $\hat{\Psi}$ based
on $922$
bootstrap runs}\label{tabpsivarcov}
\begin{tabular*}{\tablewidth}{@{\extracolsep{\fill}}lcd{3.2}d{4.2}d{3.2}
d{4.2}d{4.2}@{}}
\hline
& \multicolumn{1}{c}{$\bolds{\beta_{1}}$}
& \multicolumn{1}{c}{$\bolds{\beta_{2}}$}
& \multicolumn{1}{c}{$\bolds{\sigma_{\varepsilon}^{2}}$}
& \multicolumn{1}{c}{$\bolds{\gamma}$}
& \multicolumn{1}{c}{$\bolds{\theta}$}
& \multicolumn{1}{c@{}}{$\bolds{\phi}$} \\
\hline
$\beta_{1}$ & 303.00 &
31.87 &
207.95 & 50.04 &
108.58 & 700.60 \\
$\beta_{2}$ &  & 217.28 &
450.01 & 90.42 &
899.77 & 1091.76 \\[2pt]
$\sigma_{\varepsilon}^{2}$ &  &
& 6180.13 & 206.79 &
4343.86 & 2725.23 \\
$\gamma$ & & &
& 51.04 & 30.94 & 578.37 \\
$\theta$ & & &
& & 8941.41
& 693.87 \\
$\phi$ & & &
& & & 7268.92\\
\hline
\end{tabular*}
\end{table}

As expected, the $\bolds{\beta}$ coefficients related to the covariates
are both positive in sign.
The coefficient $\beta_{1}$ related to the spatial density of the
joint-stock companies is
characterized by a larger confidence interval with lower control limit
$-0.08$. Since the
confidence interval is an approximation based on bootstrap runs, the
covariate is retained.

\begin{table}[b]
\tablewidth=273pt
\caption{Approximated Hessian matrix for $\hat{\Psi}$} \label{tabhessian}
\begin{tabular*}{\tablewidth}{@{\extracolsep{\fill}}lcd{2.2}d{4.2}d{4.2}d{4.2}d{4.2}@{}}
\hline
& \multicolumn{1}{c}{$\bolds{\beta_{1}}$}
& \multicolumn{1}{c}{$\bolds{\beta_{2}}$}
& \multicolumn{1}{c}{$\bolds{\sigma_{\varepsilon}^{2}}$}
& \multicolumn{1}{c}{$\bolds{\gamma}$}
& \multicolumn{1}{c}{$\bolds{\theta}$}
& \multicolumn{1}{c@{}}{$\bolds{\phi}$} \\
\hline
$\beta_{1}$ & 158.83 & -2.97 &
-46.78 & 14.78 &
-23.55 & 231.79 \\
$\beta_{2}$ & & 66.68 &
-46.53 & 14.70 &
-23.42 & 230.56 \\[2pt]
$\sigma_{\varepsilon}^{2}$ & &
& 5274.86 & -207.60 &
2915.79 & -371.20 \\
$\gamma$ & & & & 14.10 &
-99.79 & 117.26 \\
$\theta$ & & &
& & 4412.62
& -186.89 \\
$\phi$ & & &
& & & 1839.45\\
\hline
\end{tabular*}
\end{table}

\begin{figure}

\includegraphics{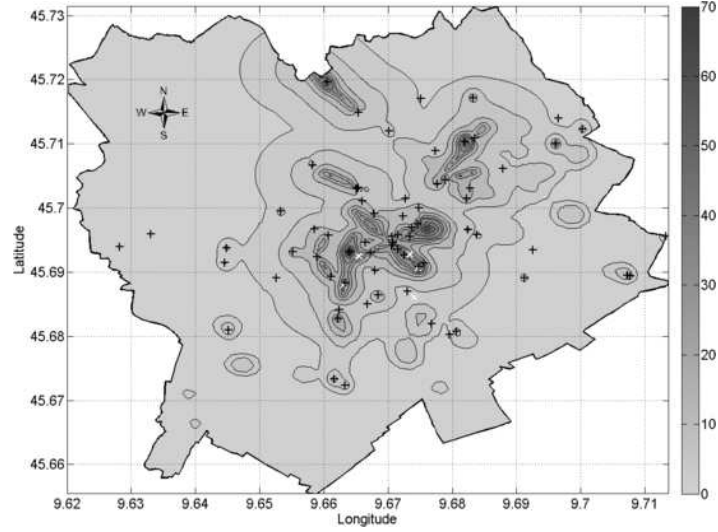}

\caption{Estimated potential $q_{\hat{\Psi}}(\mathbf{s})$ (average
daily number of copies) over the area of the city of Bergamo.}
\label{potentialline}
\end{figure}

The estimated $\hat{\theta}$ $\simeq82m$ suggests that the potential $q$,
net of the covariate, is not highly spatially correlated.
Moreover, as supported by $\hat{\phi}$ $\simeq231m$, the competition between
nearby newsstands is quite strong and two newsstands $200m$ apart measure/absorb
(on average) only 70\% of the actual market potential at their locations.

\begin{figure}

\includegraphics{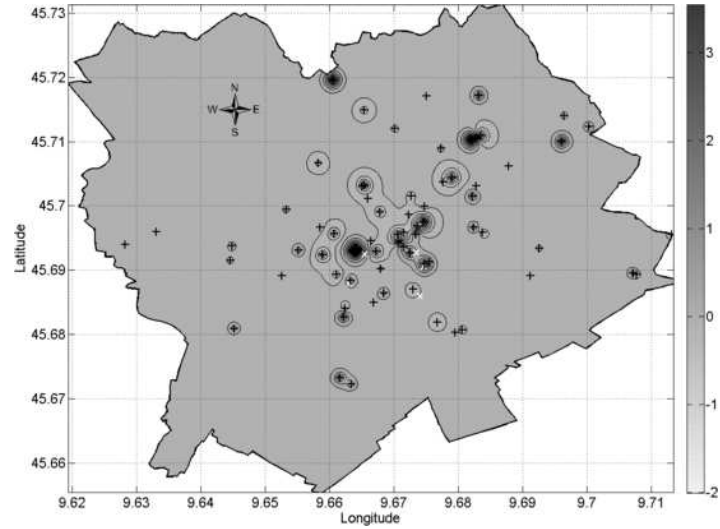}

\caption{Estimated latent variable $\hat{w}(\mathbf{s})$ over the
area of the city of Bergamo.}
\label{latentvariable}
\end{figure}

\begin{figure}[b]

\includegraphics{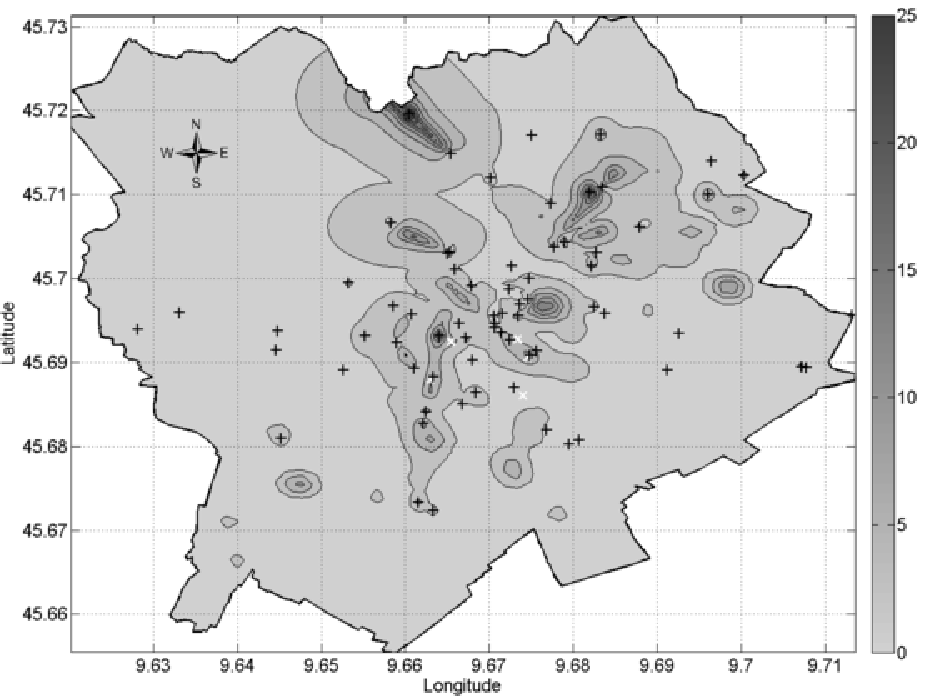}

\caption{Estimated conditional potential $q_{\hat{\Psi}}(\mathbf
{s};\mathcal{S})$ (average daily number of copies) over the area of
the city of Bergamo.}
\label{conditionalpotentialline}
\end{figure}

\begin{figure}

\includegraphics{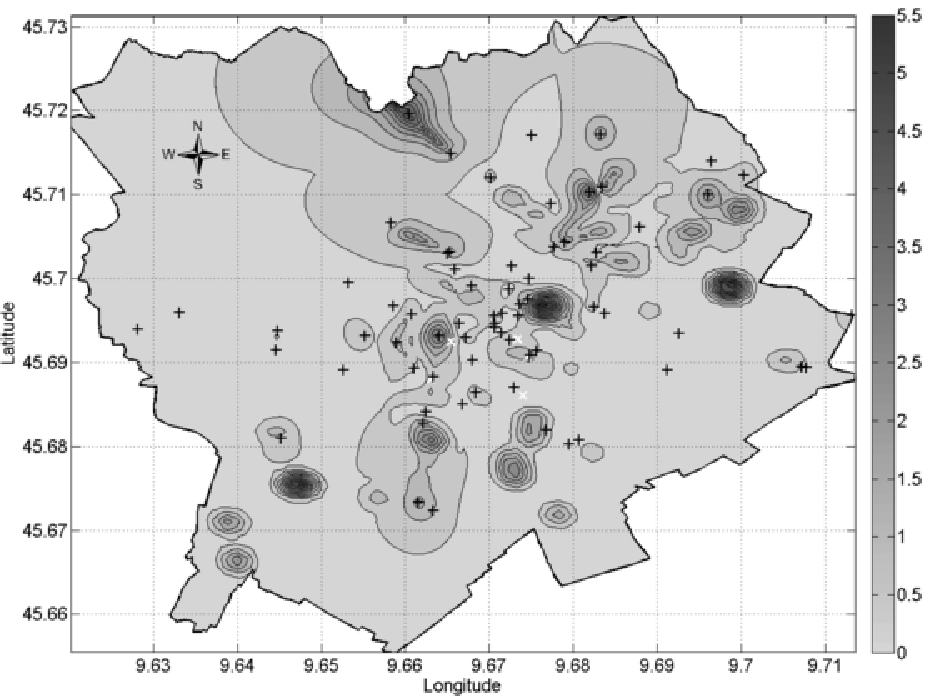}

\caption{Bootstrap standard deviation map of the estimated conditional
potential $q_{\hat{\Psi}}(\mathbf{s};\mathcal{S})$.}
\label{conditionalstd}\vspace*{3pt}
\end{figure}

Given $\hat{\Psi}$, considering equation (\ref{eqkriging}), the
market potential $q_{\hat{\Psi}}(\mathbf{s})$
is estimated over the area of the city of Bergamo as depicted in Figure
\ref{potentialline}. For each $\mathbf{s}\in\mathcal{D}$, $q_{\hat
{\Psi}}(\mathbf{s})$ provides the daily average newspaper number of
copies that would be sold by a newsstand if placed at $\mathbf{s}$
without any other newsstand in $\mathcal{D}$.
The maxima of $q_{\hat{\Psi}}(\mathbf{s})$ correspond to
commercially strategic areas that should be served by at least one
newsstand. The global
maximum is equal to $79.86$ yearly average newspaper copies and it is
located at $45.6930^{\circ}$ latitude and $9.6640^{\circ}$ longitude.
The estimated latent variable $\hat{w}(\mathbf{s})$ is displayed in
Figure~\ref{latentvariable}. It can be noted that its role is more pronounced
in the city center where, apparently, the covariates are less capable
of explaining the observed market potential.
The estimated conditional market potential $q_{\hat{\Psi}}(\mathbf
{s};\mathcal{S})$, depicted in Figure
\ref{conditionalpotentialline}, provides the daily average newspaper
number of
copies that would be sold by a newsstand if placed at $\mathbf{s}$
given the current newsstands located at $\mathcal{S}$.
Thus, the maxima of $q_{\hat{\Psi}}(\mathbf{s};\mathcal{S})$
represent the spatial locations where it would be
profitable to open additional newsstands. Finally, the bootstrap
standard deviation of the estimated conditional market potential, representing
its uncertainty, is shown in Figure~\ref{conditionalstd}.

\begin{figure}[b]\vspace*{3pt}

\includegraphics{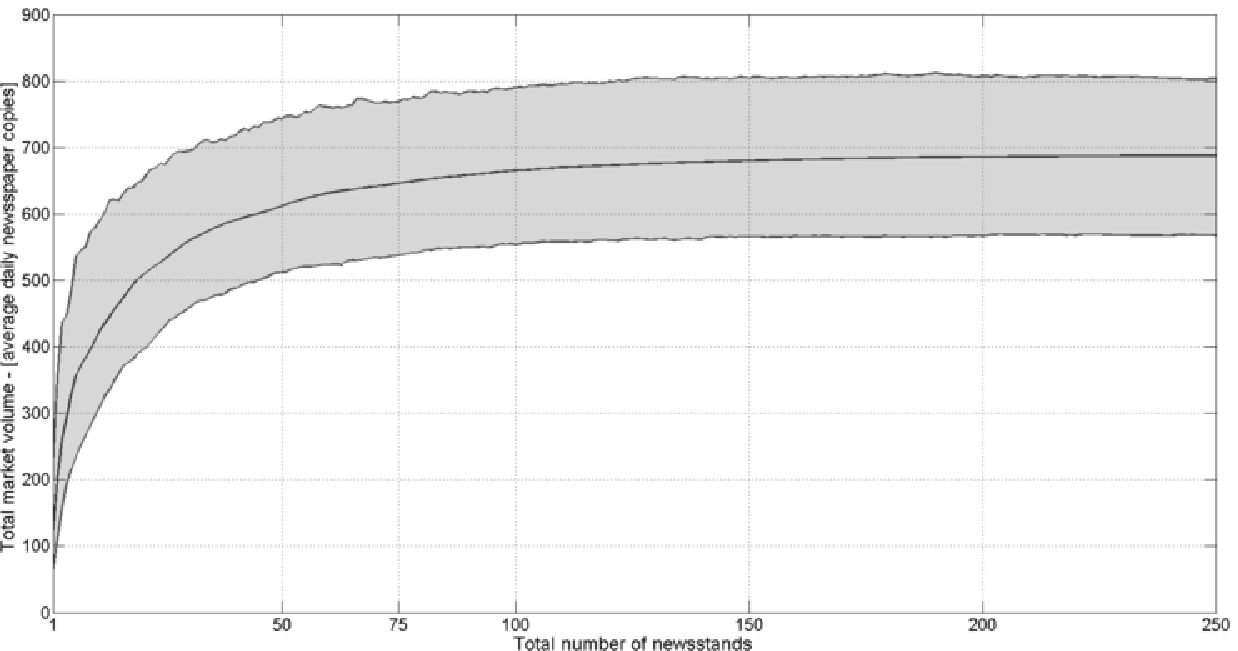}

\caption{Total market volume with respect to the number of newsstands
and $95\%$ confidence interval based on $922$ bootstrap runs.}
\label{totalmarketvolume}
\end{figure}

The total market volume related to the economic newspaper and the
Bergamo area has been evaluated following
the procedure discussed in Section~\ref{partotalpotential}. Figure
\ref{totalmarketvolume} shows the total market volume and its $95\%$
confidence interval with respect to the number of newsstands. Note that
%
the market volume stabilizes at around
$688$ average daily copies after $200$ newsstands. This is a
consequence of the fact that $200$ newsstands
absorb most of the market potential and adding more newsstands does not
increase the total market volume significantly.
Also, note that the optimized retail network of $75$ newsstands absorbs
a market volume equal to
$646$ copies, which corresponds to $94\%$ of the maximum market volume
and is $30.8\%$ higher than
the market volume absorbed by the actual retail network.

In light of this result, the publisher of the economic daily newspaper
can consider improving the retail network in order to increase the
daily market volume. Additional
newsstands can be opened only with the consent of the municipal
authority and, since the economic newspaper
represents a small part of the daily revenue of a newsstand, it cannot
be guaranteed that the additional newsstands
would be opened where the market potential of the newspaper is high.
Nevertheless, the Italian market of daily
newspapers is undergoing a liberalization phase and the publisher
should start thinking of new forms of retailing.\

\section{Conclusions}\label{secconclusions}\label{sec6}

The geostatistical potential model has been proven to be an essential
statistical tool
for the estimation of the spatial market potential of a retail product
from its sales data.
The model output is immediately and easily interpretable (uncertainty
included), as it is provided
in the form of spatially continuous surfaces and with the same unit of
measure of the original data.

The geostatistical potential model has been successfully applied to the
estimation of the spatial
market potential and to the total market volume of an economic daily
newspaper for the city of Bergamo, Italy.
The analysis of the results allows us to conclude that the daily sales
volume can be significantly increased
by focusing on the areas of the city which are characterized by a high
market potential but they are
not properly covered by the retail network.

Future extensions of the model include the introduction of the time
variable, in order to describe and
study the temporal fluctuations of the spatial market potential, and
the relaxation of the equally-effectiveness
property, in order to address the case of stores characterized by a
different degree of attractiveness.\looseness=-1

As a final remark, it is worth noting that the geostatistical potential
model can be applied outside
the geomarketing field. For instance, the sales data of the chemists of
a city with respect to a drug
can be analyzed in order to assess the diffusion of a disease in terms
of a spatially continuous surface.
More generally, the model can be applied in the case where the data
related to a set of statistical
units are available in aggregated form (e.g., due to privacy reasons)
but they are georeferenced with
respect to precise points in space.\looseness=-1

\begin{appendix}\label{app}
\section{Latent variable estimation}\label{appA}

The Gaussian latent variable $\mathbf{w}$ is estimated by applying the
usual formulas
of the multivariate normal distribution. In particular,%
%
\begin{eqnarray}\label{eqwhat}
\hat{\mathbf{w}} &=&E_{\Psi^{(k)}} ( \mathbf{w}\mid\mathbf {y} )
\nonumber
\\
&=&\bolds{\Sigma}_{\mathbf{wy}}\bolds{\Sigma}_{\mathbf{y}}^{-1} \bigl[
\mathbf{y}-E ( \mathbf{y} ) \bigr]
\\
&=&\bolds{\Sigma}_{\mathbf{wy}}\bolds{\Sigma}_{\mathbf{y}}^{-1} \bigl[
\mathbf{y}-\mathbf{G} ( \bolds{1}\mu+\mathbf{X}\bolds{\beta} ) %
\bigr],\nonumber
\end{eqnarray}
where%
\begin{eqnarray*}
\bolds{\Sigma}_{\mathbf{y}} &=&\operatorname{Var} \bigl[ \mathbf{G} ( \bolds{1}\mu+
\mathbf{X }\bolds{\beta}+\gamma\mathbf{w}+\bolds{\varepsilon}%
) \bigr]
\\
&=&\mathbf{G}\operatorname{Var} [ \gamma\mathbf{w}+\bolds{\varepsilon} ] \mathbf{G}^{\prime}
\\
&=&\mathbf{G} \bigl( \gamma^{2}\bolds{\Sigma}_{\mathbf{w}}+\bolds{
\Sigma}%
_{\bolds{\varepsilon}} \bigr) \mathbf{G}^{\prime}
\end{eqnarray*}
and%
\begin{eqnarray*}
\bolds{\Sigma}_{\mathbf{wy}} &=&E \bigl[ ( \mathbf{w}-\mathbf{0} 
)
\cdot \bigl[ \mathbf{y}-E ( \mathbf{y} ) \bigr]^{\prime}%
\bigr]
\\[-2pt]
&=&E \bigl[ \mathbf{w}\cdot ( \gamma\mathbf{Gw} )^{\prime} \bigr]
\\[-2pt]
&=&\gamma\bolds{\Sigma}_{\mathbf{w}}\mathbf{G}^{\prime}.\vadjust{\goodbreak}
\end{eqnarray*}
The variance of the estimated $\hat{\mathbf{w}}$ is given by
%
\begin{eqnarray}\label{eqAhat}
\hat{\mathbf{A}} &=&\operatorname{Var}_{\Psi^{(k)}} ( \mathbf{w}\mid\mathbf {y} )
\nonumber\\[-9pt]\\[-9pt]
&=&\bolds{\Sigma}_{\mathbf{w}}-\bolds{\Sigma}_{\mathbf{wy}}
\bolds{\Sigma}_{\mathbf{y}}^{-1} ( \bolds{\Sigma}_{\mathbf{wy}} )^{\prime}.
\nonumber
\end{eqnarray}
When $\mathbf{y}$ is characterized by missing data, equations
(\ref{eqwhat}) and (\ref{eqAhat}) become
\begin{eqnarray*}
\hat{\mathbf{w}} &=& \bigl( \bolds{\Sigma}_{\mathbf{wy}}\mathbf {L}^{\prime
}
\bigr) \bigl( \mathbf{L}\Sigma_{\mathbf{y}}\mathbf{L}^{\prime
}
\bigr)^{-1} \bigl[ \mathbf{L} \bigl( \mathbf{y}-\mathbf{G} (
\bolds{1}\mu+%
\mathbf{X }\bolds{\beta} ) \bigr) \bigr] ,
\\[-2pt]
\hat{\mathbf{A}} &=&\bolds{\Sigma}_{\mathbf{w}}- \bigl( \bolds{\Sigma
}_{%
\mathbf{wy}}\mathbf{L}^{\prime} \bigr) \bigl( \mathbf{L}
\Sigma_{\mathbf{y}}%
\mathbf{L}^{\prime} \bigr)^{-1} (
\mathbf{L}\Sigma_{\mathbf
{wy}%
} ).
\end{eqnarray*}

\section{Vector and matrix derivatives}\label{secB}\label{appB}

The evaluation of the approximate Fisher information matrix defined in
equation (\ref{eqinformationmatrix}) requires the computation of the vector
derivatives $%
\partial\bolds{\epsilon} ( \Psi ) /\partial\Psi_{i}$ and the
matrix derivatives $\partial\bolds{\Sigma}_{\bolds{\epsilon}} ( \Psi )
/\partial\Psi_{i}$, $1\leq i\leq\llvert \Psi \rrvert $, with
$\bolds{\epsilon}$ and $\bolds{\Sigma}_{\bolds{\epsilon}}$ defined in equations
(\ref{eqinnovation}) and (\ref{eqinnovationvariance}), respectively.

In the case of the spatial correlation function defined in equation
(\ref{eqrhosim})
and the interaction function defined in equation (\ref{eqfsim}), the
following derivatives hold:
%
\begin{eqnarray}
\frac{\partial\bolds{\epsilon} ( \Psi ) }{\partial\Psi_{i}} &=&
\cases{ -\mathbf{g}, &\quad if $\Psi_{i}=\mu$,
\vspace*{1pt}\cr
-
\mathbf{g}\odot\mathbf{x}_{l}, &\quad if $\Psi_{i}=
\beta_{l}; 1\leq l\leq b$,
\vspace*{1pt}\cr
-\partial\mathbf{g}_{\phi}\odot
( \bolds{1}\mu+\mathbf {X}\beta%
), &\quad if $\Psi_{i}=
\phi$,
\vspace*{1pt}\cr
\mathbf{0}, &\quad otherwise,}
\\[-2pt]
\frac{\partial\bolds{\Sigma}_{\bolds{\epsilon}} ( \Psi )
}{\partial
\Psi_{i}} &=&\cases{ \mathbf{gg}^{\prime}\odot I_{N}, &\quad
if $\Psi_{i}={\small\sigma}%
_{\varepsilon}^{2}$,
\vspace*{1pt}\cr
2\gamma\mathbf{gg}^{\prime}\odot\bolds{\Sigma}_{\mathbf{w}}, &\quad if $
\Psi_{i}=\gamma$,
\vspace*{2pt}\cr
\displaystyle \gamma^{2}\mathbf{gg}^{\prime}
\odot\frac{\mathbf{H}}{\theta^{2}}\odot%
\bolds{\Sigma}_{\mathbf{w}}, &\quad if $
\Psi_{i}=\theta$,
\vspace*{2pt}\cr
\tilde{\mathbf{G}}\odot \bigl(
\gamma^{2}\bolds{\Sigma}_{\mathbf
{w}}+%
\bolds{\Sigma}_{\bolds{\varepsilon}} \bigr), &\quad if $\Psi_{i}=\phi$,
\vspace*{1pt}\cr
\mathbf{0}, &\quad
otherwise,}
\end{eqnarray}
where $\mathbf{x}_{l}$ is the $l$th column of the matrix $\mathbf
{X}$ and $%
\mathbf{H}$ is the distance matrix based on~$\mathcal{S}$.

Finally, the $(pq)$th element of the matrix $\tilde{\mathbf{G}}$ is given
by $\partial g_{p}\cdot g_{q}+g_{p}\partial g_{q}$, where $g_{p}$ is
the $p$th element of the vector $\mathbf{g}$ and while the $p$th element
of the
vector $\partial\mathbf{g}_{\phi}$ is given by%
%
\begin{equation}
\partial g_{p}\equiv\frac{\partial g_{p}}{\partial\phi}=-\frac{%
\sum_{q\neq p}^{N}({h_{pq}}/{\phi^{2}})\exp ( -
{\llVert
\mathbf{s}_{p}-\mathbf{s}_{q}\rrVert }/{\phi} ) }{ [
\sum_{q\neq p}^{N}\exp ( -{\llVert \mathbf
{s}_{p}-\mathbf{s%
}_{q}\rrVert }/{\phi} )  ]^{2}}
\end{equation}
with $h_{pq}$ the $(pq)$th element of the matrix $\mathbf{H}$.\vadjust{\goodbreak}
\end{appendix}

\section*{Acknowledgments}

Special thanks to Alberto Saccardi of Nunatac Srl for providing the
data, to Ilaria Cremonesi for the preliminary data analysis and to
Alessandro Fass\`{o} whose advice is always precious.

\begin{supplement}
\stitle{Data set and Matlab\textregistered\ code}
\slink[doi]{10.1214/12-AOAS588SUPP}  
\sdatatype{.zip}
\sfilename{aoas588\_supp.zip}
\sdescription{Georeferentiated newsstand sales data and
Matlab\textregistered\ code for the data analysis.}
\end{supplement}


\printaddresses


\begin{thebibliography}{18}

\bibitem[\protect\citeauthoryear{Banerjee, Gelfand and
  Polasek}{2000}]{Banerjee2000}
\begin{barticle}[mr]
\bauthor{\bsnm{Banerjee},~\bfnm{Sudipto}\binits{S.}},
  \bauthor{\bsnm{Gelfand},~\bfnm{Alan~E.}\binits{A.~E.}} \AND
  \bauthor{\bsnm{Polasek},~\bfnm{Wolfgang}\binits{W.}}
(\byear{2000}).
\btitle{Geostatistical modelling for spatial interaction data with application
  to postal service performance}.
\bjournal{J. Statist. Plann. Inference}
\bvolume{90}
\bpages{87--105}.
\bid{doi={10.1016/S0378-3758(00)00111-7}, issn={0378-3758}, mr={1791583}}
\bptok{imsref}%
\end{barticle}
\endbibitem

\bibitem[\protect\citeauthoryear{Cliquet}{2006}]{cliquet2006}
\begin{bbook}[author]
\bauthor{\bsnm{Cliquet},~\bfnm{G.}\binits{G.}}
(\byear{2006}).
\btitle{Geomarketing: Methods and Strategies in Spatial Martketing}.
\bpublisher{ISTE}, \blocation{London}.
\bptok{imsref}%
\end{bbook}
\endbibitem

\bibitem[\protect\citeauthoryear{Cressie and Wikle}{2011}]{cressiewikle2011}
\begin{bbook}[mr]
\bauthor{\bsnm{Cressie},~\bfnm{Noel}\binits{N.}} \AND
  \bauthor{\bsnm{Wikle},~\bfnm{Christopher~K.}\binits{C.~K.}}
(\byear{2011}).
\btitle{Statistics for Spatio-Temporal Data}.
\bpublisher{Wiley}, \blocation{Hoboken, NJ}.
\bid{mr={2848400}}
\bptok{imsref}%
\end{bbook}
\endbibitem

\bibitem[\protect\citeauthoryear{Davis}{2006}]{Davis2006}
\begin{barticle}[author]
\bauthor{\bsnm{Davis},~\bfnm{Peter}\binits{P.}}
(\byear{2006}).
\btitle{Spatial competition in retail markets: Movie theaters}.
\bjournal{The RAND Journal of Economics}
\bvolume{37}
\bpages{964--982}.
\bptok{imsref}%
\end{barticle}
\endbibitem

\bibitem[\protect\citeauthoryear{de~Grange, Ibeas and
  Gonzalez}{2011}]{Grange2009}
\begin{barticle}[author]
\bauthor{\bparticle{de} \bsnm{Grange},~\bfnm{Louis}\binits{L.}},
  \bauthor{\bsnm{Ibeas},~\bfnm{Angel}\binits{A.}} \AND
  \bauthor{\bsnm{Gonzalez},~\bfnm{Felipe}\binits{F.}}
(\byear{2011}).
\btitle{A hierarchical gravity model with spatial correlation: Mathematical
  formulation and parameter estimation}.
\bjournal{Netw. Spat. Econ.}
\bvolume{11}
\bpages{439--463}.
\bptok{imsref}%
\end{barticle}
\endbibitem

\bibitem[\protect\citeauthoryear{Diggle, Menezes and Su}{2010}]{Diggle2010}
\begin{barticle}[mr]
\bauthor{\bsnm{Diggle},~\bfnm{Peter~J.}\binits{P.~J.}},
  \bauthor{\bsnm{Menezes},~\bfnm{Raquel}\binits{R.}} \AND
  \bauthor{\bsnm{Su},~\bfnm{Ting-li}\binits{T.-l.}}
(\byear{2010}).
\btitle{Geostatistical inference under preferential sampling}.
\bjournal{J. R. Stat. Soc. Ser. C. Appl. Stat.}
\bvolume{59}
\bpages{191--232}.
\bid{doi={10.1111/j.1467-9876.2009.00701.x}, issn={0035-9254}, mr={2744471}}
\bptok{imsref}%
\end{barticle}
\endbibitem

\bibitem[\protect\citeauthoryear{Diggle and Ribeiro}{2007}]{Diggle2007}
\begin{bbook}[mr]
\bauthor{\bsnm{Diggle},~\bfnm{Peter~J.}\binits{P.~J.}} \AND
  \bauthor{\bsnm{Ribeiro},~\bfnm{Paulo~J.}\binits{P.~J.} \bsuffix{Jr.}}
(\byear{2007}).
\btitle{Model-Based Geostatistics}.
\bpublisher{Springer}, \blocation{New York}.
\bid{mr={2293378}}
\bptok{imsref}%
\end{bbook}
\endbibitem

\bibitem[\protect\citeauthoryear{Fass{\`{o}} and
  Cameletti}{2010}]{Fassocameletti2010}
\begin{barticle}[author]
\bauthor{\bsnm{Fass{\`{o}}},~\bfnm{A.}\binits{A.}} \AND
  \bauthor{\bsnm{Cameletti},~\bfnm{M.}\binits{M.}}
(\byear{2010}).
\btitle{A unified statistical approach for simulation, modeling, analysis and
  mapping of environmental data}.
\bjournal{Simulation}
\bvolume{86}
\bpages{139--154}.
\bptok{imsref}%
\end{barticle}
\endbibitem

\bibitem[\protect\citeauthoryear{Fass{\`o} and
  Finazzi}{2011}]{Fassofinazzi2011}
\begin{barticle}[mr]
\bauthor{\bsnm{Fass{\`o}},~\bfnm{Alessandro}\binits{A.}} \AND
  \bauthor{\bsnm{Finazzi},~\bfnm{Francesco}\binits{F.}}
(\byear{2011}).
\btitle{Maximum likelihood estimation of the dynamic coregionalization model
  with heterotopic data}.
\bjournal{Environmetrics}
\bvolume{22}
\bpages{735--748}.
\bid{doi={10.1002/env.1123}, issn={1180-4009}, mr={2843140}}
\bptok{imsref}%
\end{barticle}
\endbibitem

\bibitem[\protect\citeauthoryear{Finazzi}{2013}]{Finazzi2012}
\begin{bmisc}[author]
\bauthor{\bsnm{Finazzi},~\bfnm{Francesco}\binits{F.}}
(\byear{2013}).
\btitle{Supplement to ``Geostatistical modeling in the presence of interaction between
the measuring instruments, with an application to the estimation of
spatial market potentials.'' DOI:\doiurl{10.1214/12-AOAS588SUPP}.}
\bptok{imsref}%
\end{bmisc}
\endbibitem

\bibitem[\protect\citeauthoryear{Fischer and Wang}{2011}]{Manfred2011}
\begin{bincollection}[author]
\bauthor{\bsnm{Fischer},~\bfnm{Manfred~M.}\binits{M.~M.}} \AND
  \bauthor{\bsnm{Wang},~\bfnm{Jinfeng}\binits{J.}}
(\byear{2011}).
\btitle{Spatial interaction models and spatial dependence}.
In \bbooktitle{Spatial Data Analysis}
\bpages{61--70}.
\bpublisher{Springer}, \blocation{Berlin}.
\bptok{imsref}%
\end{bincollection}
\endbibitem

\bibitem[\protect\citeauthoryear{Huff}{1964}]{Huff1964}
\begin{barticle}[author]
\bauthor{\bsnm{Huff},~\bfnm{David~L}\binits{D.~L.}}
(\byear{1964}).
\btitle{Defining and estimating a trading area}.
\bjournal{The Journal of Marketing}
\bvolume{28}
\bpages{34--38}.
\bptok{imsref}%
\end{barticle}
\endbibitem

\bibitem[\protect\citeauthoryear{McLachlan and Krishnan}{2008}]{McLachlan2008}
\begin{bbook}[mr]
\bauthor{\bsnm{McLachlan},~\bfnm{Geoffrey~J.}\binits{G.~J.}} \AND
  \bauthor{\bsnm{Krishnan},~\bfnm{Thriyambakam}\binits{T.}}
(\byear{2008}).
\btitle{The {EM} Algorithm and Extensions},
\bedition{2nd} ed.
\bpublisher{Wiley}, \blocation{Hoboken, NJ}.
\bid{doi={10.1002/9780470191613}, mr={2392878}}
\bptok{imsref}%
\end{bbook}
\endbibitem

\bibitem[\protect\citeauthoryear{Meng and Rubin}{1991}]{Mengrubin1991}
\begin{barticle}[author]
\bauthor{\bsnm{Meng},~\bfnm{Xiao~L.}\binits{X.~L.}} \AND
  \bauthor{\bsnm{Rubin},~\bfnm{Donald~B.}\binits{D.~B.}}
(\byear{1991}).
\btitle{Using {EM} to {o}btain {a}symptotic {v}ariance--{c}ovariance matrices:
{T}he {SEM} algorithm}.
\bjournal{J. Amer. Statist. Assoc.}
\bvolume{86}
\bpages{899--909}.
\bptok{imsref}%
\end{barticle}
\endbibitem

\bibitem[\protect\citeauthoryear{Reilly}{1931}]{reilly1931}
\begin{bbook}[author]
\bauthor{\bsnm{Reilly},~\bfnm{W.~J.}\binits{W.~J.}}
(\byear{1931}).
\btitle{The Law of Retail Gravitation}.
\bpublisher{Reilly}, \blocation{New York}.
\bptok{imsref}%
\end{bbook}
\endbibitem

\bibitem[\protect\citeauthoryear{Shumway and
  Stoffer}{2006}]{Shumwaystoffer2006}
\begin{bbook}[mr]
\bauthor{\bsnm{Shumway},~\bfnm{Robert~H.}\binits{R.~H.}} \AND
  \bauthor{\bsnm{Stoffer},~\bfnm{David~S.}\binits{D.~S.}}
(\byear{2006}).
\btitle{Time Series Analysis and Its Applications: With R Examples},
\bedition{2nd} ed.
\bpublisher{Springer}, \blocation{New York}.
\bid{mr={2228626}}
\bptok{imsref}%
\end{bbook}
\endbibitem

\bibitem[\protect\citeauthoryear{Xu and Wikle}{2007}]{XuWikle2007}
\begin{barticle}[mr]
\bauthor{\bsnm{Xu},~\bfnm{Ke}\binits{K.}} \AND
  \bauthor{\bsnm{Wikle},~\bfnm{Christopher~K.}\binits{C.~K.}}
(\byear{2007}).
\btitle{Estimation of parameterized spatio-temporal dynamic models}.
\bjournal{J.~Statist. Plann. Inference}
\bvolume{137}
\bpages{567--588}.
\bid{doi={10.1016/j.jspi.2005.12.005}, issn={0378-3758}, mr={2298958}}
\bptok{imsref}%
\end{barticle}
\endbibitem

\end{thebibliography}
\end{document}